\documentclass[final]{siamltex}
\usepackage{amsmath}
\usepackage{amsfonts}
\usepackage{amssymb,amsxtra}
\usepackage{graphicx}

\def\centereps#1#2#3{\vskip#2\relax\centerline{\hbox to#1{\special
  {eps:#3 x=#1, y=#2}\hfil}}}

\title{Deflated and Restarted Symmetric Lanczos Methods for Eigenvalues and Linear Equations with Multiple Right-hand Sides \footnotemark[1]}

\author{Abdou M. Abdel-Rehim \footnotemark[2]
\and Ronald B. Morgan\footnotemark[3]
\and Dywayne A. Nicely\footnotemark[4]
\and Walter Wilcox\footnotemark[5] }

\begin{document}
\bibliographystyle{plain}

\maketitle

\renewcommand{\thefootnote}{\fnsymbol{footnote}}
\footnotetext[1]{This work was partially supported by the National Science Foundation, Computational Mathematics Program under grant 0310573 and the National Computational Science Alliance.  It utilized the Baylor High-Performance Computing Cluster.  The second author was also supported by the Baylor University Sabbatical Program.}
\footnotetext[2]{Department of Physics, Baylor
University, Waco, TX 76798-7316.  ({\tt Abdou\_Abdel-Rehim@baylor.edu}).}
\footnotetext[3]{Department of Mathematics, Baylor
University, Waco, TX 76798-7328 ({\tt Ronald\_Morgan@baylor.edu}).}
\footnotetext[4]{Department of Mathematics, Baylor
University, Waco, TX 76798-7328 ({\tt Dywayne\_Nicely@baylor.edu}).}
\footnotetext[5]{Department of Physics, Baylor
University, Waco, TX 76798-7316.  ({\tt Walter\_Wilcox@baylor.edu}).}
\renewcommand{\thefootnote}{\arabic{footnote}}

\begin{abstract}

A deflated restarted Lanczos algorithm is given for both solving symmetric linear equations and computing eigenvalues and eigenvectors.  The restarting limits the storage so that finding eigenvectors is practical.  Meanwhile, the deflating from the presence of the eigenvectors allows the linear equations to generally have good convergence in spite of the restarting.  Some reorthogonalization is necessary to control roundoff error, and several approaches are discussed.  The eigenvectors generated while solving the linear equations can be used to help solve systems with multiple right-hand sides.  Experiments are given with large matrices from quantum chromodynamics that have many right-hand sides.  

\end{abstract}



\pagestyle{myheadings}
\thispagestyle{plain}
\markboth{A. M. ABDEL-REHIM, R. B. MORGAN, D. A. NICELY and W. WILCOX}{DEFLATED LANCZOS}

\section{Introduction}

We consider a large matrix $A$ that is either real symmetric or complex Hermitian.  We are interested in solving the system of equations $Ax = b$, possibly with multiple right-hand sides, and in solving the associated eigenvalue problem.  Both eigenvalues and eigenvectors are desired.  Symmetric and Hermitian problems can take advantage of fast algorithms such as the conjugate gradient method (CG)~\cite{HeSt,Sa96} for linear equations and the related Lanczos algorithm~\cite{La50,Pa} for eigenvalues.  However, regular CG can be improved upon for the case of multiple right-hand sides, and Lanczos may have storage and accuracy issues while computing eigenvectors.  We give new methods for these problems.  

An approach is presented called Lanczos with deflated restarting or Lan-DR.  It simultaneously solves the linear equations and computes the eigenvalues and eigenvectors.  A restarted Krylov subspace approach is used for the linear equations, but it also saves approximate eigenvectors at the restart and uses them in the subspace for the next cycle.  The restarting of the Lanczos algorithm does not slow down the convergence as it normally would, because once the approximate eigenvectors are accurate enough, they essentially remove or deflate the associated eigenvalues from the problem.  The eigenvalue portion of Lan-DR has already been presented in~\cite{WuSi}, but as mentioned, we add on the solution of linear equations.  Also, some reorthogonalization is necessary to control roundoff error.  We give some new approaches for this reorthogonalization.  We also give a Minres/harmonic version of the algorithm.  

An important example where both linear equations need to be solved and eigenvectors are desired is the case of multiple right-hand sides.  The eigenvector information generated for one right-hand side can be used to improve the convergence of the systems with subsequent right-hand sides.  We give a method called deflated conjugate gradients or D-CG that implements such an approach.  For the second and subsequent right-hand sides, there is first a projection over the approximate eigenvectors generated by Lan-DR on the first right-hand side.  Then the conjugate gradient iteration is applied.
We will give application to large Hermitian systems of linear equations in quantum chromodynamics (QCD).  

Section 2 has a review of methods that this paper builds on.  The Lan-DR method for eigenvalues and linear equations is presented in Section 3.  Section 4 has comparison of reorthogonalization approaches.  Multiple right-hand sides are considered in Section 5.  Then Section 6 has Minres-DR and deflated Minres, which are of interest particularly for  indefinite systems and interior eigenvalues.

\section{Review}

\subsection{Restarted methods for eigenvalue problems}

Restarted Krylov methods for nonsymmetric eigenvalue problems took a jump forward with the implicitly restarted Arnoldi method (IRAM) by Sorensen~\cite{So}.  IRAM restarts with several vectors, so it not only computes many eigenvectors simultaneously, but it also has improved convergence.  At the time of a restart, let the Ritz vectors be $\{y_1, y_2, \ldots, y_k\}$ and let $r$ be a multiple of the residual vectors for these Ritz vectors (the residuals are parallel).  Then the next cycle of IRAM builds the subspace 
\begin{equation}
Span\{y_1, y_2, \ldots y_k, r, A r, A^2 r, A^3 r, \ldots ,A^{m-k-1} r \}. \label{ss}
\end{equation}
This subspace is equivalent~\cite{Arnoldi-R} to 
\begin{equation}
Span\{y_1, y_2, \ldots y_k, A y_i, A^2 y_i, A^3 y_i, \ldots ,A^{m-k} y_i \}. \label{ss2}
\end{equation}
This last form helps show why IRAM is effective.  The subspace contains a Krylov subspace with each approximate eigenvector as starting vector.  

Versions of IRAM for symmetric problems are given in~\cite{CaReSo,BaCaRe96}.  
A mathematically equivalent method to IRAM that does not use the implicit restarting is in~\cite{Arnoldi-R} (this approach is equivalent at the end of each cycle).
Wu and Simon present a mathematically equivalent approach for symmetric problems called thick restarted Lanczos (TRLAN)~\cite{WuSi}.  They put the approximate eigenvectors at the beginning of a new subspace instead of at the end as was done in~\cite{Arnoldi-R}.  Stewart gives a framework for restarted Krylov methods in~\cite{St01}.  See~\cite{Sa04} for a symmetric block version.  Nonsymmetric and harmonic versions of restarted Arnoldi following the TRLAN approach are in~\cite{HRAM}.  For block Arnoldi methods, see~\cite{Ba07} and its references.  

\subsection{Deflated restarted methods for linear equations} 

Deflated Krylov methods for linear equations compute eigenvectors and use them to deflate eigenvalues and thus improve convergence of the linear equations solution.  For problems that have slow convergence due to small eigenvalues, deflation can make a big difference.  For nonsymmetric problems, approximate eigenvectors are formed during GMRES~\cite{SaSc} in~\cite{GMRES-E,KhYe,ErBuPo,ChSa,Sa95B,BaCaGoRe,BuEr,LCMo,DS99,GMRES-IR, GMRES-DR}.  Methods in~\cite{DS99} are related in that they save information during restarted GMRES in order to improve convergence.  

Now for symmetric problems, it is assumed in~\cite{Ni} that there is a way to get approximate eigenvectors which are used for a deflated CG.  Eigenvectors are formed during CG in~\cite{SaYeErGu}.  

We now look at a particular deflated GMRES method that will be used in this paper.  GMRES-DR(m,k)~\cite{GMRES-DR} uses the subspace 
\begin{equation}
Span\{\tilde y_1, \tilde y_2, \ldots \tilde y_k, r_0, A r_0, A^2 r_0, A^3
r_0, \ldots ,A^{m-k-1} r_0 \}, \label{ss3}
\end{equation}
where $r_0$ is the initial residual for the linear equations
at the start of the new cycle, 
$\{\tilde y_1, \tilde y_2, \ldots \tilde y_k\}$
are the harmonic Ritz vectors corresponding to the smallest harmonic Ritz values.  The dimension of the whole subspace is $m$, including the $k$ approximate eigenvectors (the harmonic Ritz vectors).
This can be viewed as a Krylov subspace generated with starting vector
$r_0$ augmented with approximate eigenvectors.  Remarkably, the whole
subspace turns out to be a Krylov subspace itself (though not with $r_0$ as
starting vector)~\cite{GMRES-IR}.  FOM-DR~\cite{GMRES-DR} is a version that, as in FOM~\cite{Sa81}, uses a Galerkin projection instead of minimum residual projection.  FOM-DR also needs regular Ritz vectors instead of harmonic.

\subsection{Multiple right-hand sides} 

Systems with multiple right-hand sides occur in many applications (see~\cite{FrMa} for some examples).  Block methods are a standard way to solve systems with multiple right-hand sides (see for example~\cite{OL80, Sa96, FrMa, bgdr, Gu07}).  However, block methods are not ideal for every circumstance.  
Other approaches for multiple right-hand sides use information from the solution of the first right-hand side (and possibly others) to assist subsequent right-hand sides.  Seed methods~\cite{SmPeMi,ChWa,KiMiRa,Pa80B,Sa87,vdV87,ErGu} project over entire subspaces generated while solving previous right-hand sides.   Simoncini and Gallopoulos~\cite{SiGa,SiGa96} suggest methods including using seeding, using blocks and using Richardson iteration with a polynomial generated from GMRES applied to the first right-hand side.   
In~\cite{gproj} a small subspace is generated with GMRES-DR applied to the first right-hand side that contains important information of approximate eigenvectors, and this is used to improve the subsequent right-hand sides.
See~\cite{PadeStMaJoMa} for a method for multiple right-hand sides that can also handle a changing matrix.

\section{Lanczos with deflated restarting}

We propose a restarted Lanczos method that both solves linear equations and computes eigenvalues and eigenvectors. It is called Lanczos with deflated restarting or Lan-DR.  The Lan-DR method is a version of FOM-DR~\cite{GMRES-DR} for symmetric and Hermitian problems and is closely related to GMRES-DR~\cite{GMRES-DR}.  As mentioned earlier, the eigenvalue portion of Lan-DR is TRLAN~\cite{WuSi} and is mathematically equivalent to implicitly restarted Arnoldi~\cite{So}.  

For Lan-DR, the number of desired eigenvectors $k$ must be chosen, along with which eigenvalues are to be targeted.  Normally the eigenvalues nearest the origin are the most important ones for deflation purposes, but other eigenpairs can be computed.  In particular, deflating large outstanding eigenvalues may help convergence of the linear equations solution and may be useful for stability.

At the time of a restart, let $r_0$ be the residual vector for the linear equations and let the Ritz vectors from the previous cycle be $\{y_1, y_2, \ldots, y_k\}$.  Then the next cycle of Lan-DR builds the subspace 
\begin{equation}
Span\{y_1, y_2, \ldots y_k, r_0, A r_0, A^2 r_0, A^3 r_0 \ldots ,A^{m-k-1} r_0 \}. \label{landrss}
\end{equation}
Lan-DR generates the recurrence formula
\begin{equation}
 AV_m = V_{m+1} \overline T_m, \label{recur0}
\end{equation} 
where $V_m$ is an $n$ by $m$ matrix whose columns span the subspace (\ref{landrss}) and $V_{m+1}$ is the same except for an extra column.  Also $\overline T_m$ is an $m+1$ by $m$ matrix that is tridiagonal except for the $k+1$ by $k+1$ leading portion.  This portion is non-zero only on the main diagonal (which has the Ritz values) and in the $k+1$ row and column.  A part of recurrence (\ref{recur0}) can be separated out to give
\begin{equation}
 AV_k = V_{k+1} \overline T_k, \label{recur1}
\end{equation} 
where $V_k$ is an $n$ by $k$ matrix whose columns span the subspace of
Ritz vectors, $V_{k+1}$ is the same except for an extra column
and $\overline T_k$ is the leading $k+1$ by $k$ portion of $T_m$.   This recurrence allows access to both the approximate eigenvectors (the Ritz vectors) and their products with $A$ while requiring
storage of only $k+1$ vectors of length $n$.  The approximate eigenvectors in Lan-DR span a small Krylov subspace of dimension $k$~\cite{GMRES-DR}. 

It is necessary to maintain some degree of orthogonality of the columns of $V_{m+1}$.  We mention here an approach to reorthogonalization that we call k-selective reorthogonalization (k-SO).  For cycles after the first one, all new Lanczos vectors are reorthogonalized against the $k$ Ritz vectors.  This uses Parlett and Scott's idea of selective reorthogonalization~\cite{PaSc}, but is more natural in this setting, because we are already computing the approximate eigenvectors.  Also, because of the restarting, there is no need to store a large subspace.  Simon's partial reorthogonalization~\cite{Si84,WuSi} is also a possibility, as is periodic reorthogonalization~\cite{Gr81}.  See Section 4 for more on this, including hybrid approaches.  
Full reorthogonalization or partial reorthogonalization can be used for the first cycle, but this may not be necessary.  By Paige's theorem~\cite{Paige71,Pa,Stv2}, orthogonality is only lost when some eigenvectors begin to converge.  This may not happen in the first cycle if there are no outstanding eigenvalues.  In our tests, we do reorthogonalize for the first cycle.

\vspace{.10in}
\begin{center}
\textbf{Lan-DR(m,k)}
\end{center}
\begin{itemize}
\item[1.]  \textit{Start.}  Choose $m,$ the maximum size of the
subspace, and $k$, the desired number of approximate eigenvectors.   
If there is an initial guess, $x_0$, then the problem becomes $A(x-x_0) = r_0$.

\item[2.]  \textit{First cycle.} Apply $m$ iterations of the standard symmetric Lanczos
algorithm. This computes the matrix $V_{m+1}$ that has the Lanczos vectors as columns and the $m+1$ by $m$ tridiagonal matrix $\overline{T}_{m}$.  In addition, fully
reorthogonalize all the Lanczos vectors.  

\item[3.] \textit{Eigenvector computation.}  Compute the $k$
smallest (or others, if desired) eigenpairs, $(\theta_{i},g_{i})$,
of $T_{m}$, the $m$ by $m$ portion of $\overline{T}_{m}$.  For $i = 1,...,k,$ form $y_{i} = V_m g_{i}$.  The shortcut residual norm formula is $r_{i}=|t_{m+1,m} g_{m,i}|$.

\item[4.] \textit{Linear equations.}  Let $c_m = V_k r_0$ (for the first cycle, this is all zeros except for $||r_0||$ in the first position, for other cycles it is all zeros except for $||r_0||$ in the $k+1$ position).  Solve $T_m d = c$, and set $\tilde x = x_0 + V_m d$.    Then $r = r_0 - A \tilde x = r_0 - V_{m+1} \overline{T}_{m} d$. If satisfied with convergence of the linear equations and the eigenvalues, can stop.  If not, let the new $x_0 = \tilde x$ and $r_0 = r$ and continue.   

\item[5.] \textit{Restart.} For $i = 1,...,k$, reassign $v_{i} = y_{i}$.  Set $v_{k+1}$ to be the previous $v_{m+1}$.  Set the $k$ by $k$ portion of $\overline{T}_{k}^{new}$ to be the diagonal matrix with the Ritz values $\theta_{i}$ as diagonal entries.  For $i = 1,...,k,$, the $i$th element $k$th row of $\overline{T}_{k}^{new}$ is computed by $t_{m+1,m} g_{m,i}$.  Set $\overline{T}_{k} = \overline{T}_{k}^{new}$. 

\item[6.]  \textit{Main iteration.} First we compute the $v_{k+2}$ vector.  Compute $w = A v_{k+1} - \sum_{i=1}^{k+1} t_{k+1,i} v_{i}$, then $t_{m+1,m+1} = v_{k+1}^T w$ and $w = w - t_{m+1,m+1} v_{k+1}$.  Let $t_{m+1,m+1} = ||w||$.  Set $v_{k+2} = w/||w||$.  Next, compute  the other $v_i$ vectors, for $i = k+3,...,m+1$, using the standard Lanczos iteration.  $\overline{T}_{m}$ is formed at the same time.  Also reorthogonalize the $v_i$ vectors as desired (see next section).  Go to step 3.
 
\end{itemize}
\vspace{.15in}


We give the expense for Lan-DR(m,k) without reorthogonalization (see the next section for reorthogonalization expenses).  The cost is $m-k$ matrix-vector products per cycle and about $6(m-k)+(k+2)m$ length $n$ vector operations (such as dot products and daxpy's) per cycle.  For $k$ small relative to $m$, Lan-DR uses one matrix-vector product and roughly $k+8$ vector ops per iteration.  For $k$ near $m/2$, there are about $2k+10$ vector ops per iteration.  
This compares to about $5$ length $n$ vector ops per iteration for CG.  Of course, CG does not calculate eigenvectors like Lan-DR does.

{\it Example 1.} We use a test matrix that has many small eigenvalues.  It is a 
diagonal matrix of dimension $n = 5000$ whose diagonal elements are $0.1, 0.2, 0.3,\ldots 9.8, 9.9, 10,$ $11, 12, \ldots ,4909, 4910.$  The right-hand side is a random normal vector.  We apply the method Lan-DR(100,40), which at each restart keeps the 40 Ritz vectors corresponding to the smallest Ritz values and then builds a subspace of dimension 100 (including the 40 approximate eigenvectors).  With k-SO, at every iteration of all of the cycles except the first, there is a reorthogonalization against 40 vectors.  We first look at how well Lan-DR computes the eigenvalues.  Figure 3.1 shows the residual norms for the 40 Ritz vectors.  The desired number of eigenvalues is 30 and the desired residual tolerance is $10^{-8}$.  It takes 57 cycles for the first 30 eigenvalues to reach this level.  Since this is a fairly difficult problem with eigenvalues clustered together, it takes a while for eigenvalues to start converging.  However, from that point, eigenvalues converge regularly, and it can be seen that many eigenvalues and their eigenvectors can be computed accurately.  The orthogonalization costs are significantly less than for fully reorthogonalized IRAM (about 84 vector operations for orthogonalization per Lan-DR iteration versus an average of 280 for IRAM).  For a matrix that is fairly sparse so that the matrix-vector product is inexpensive (and also with cheap preconditioner, if there is one), the difference in orthogonalization is significant.

\begin{figure}
\includegraphics[width=4in]{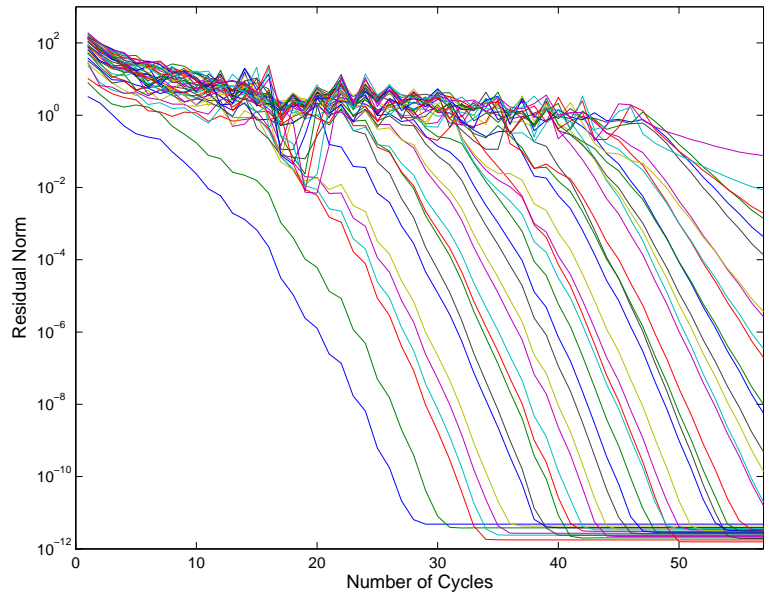}
\caption{Computing many eigenvalues of a matrix with small eigenvalues.}
\end{figure}

We continue the example by comparing Lan-DR(100,40) to unrestarted Lanczos.  Figure 3.2 has the residual norms for the smallest and 30th eigenvalues with each method.  The results are very similar for the first eigenvalue, in spite of the fact that Lan-DR is restarted.  The presence of the approximate eigenvectors corresponding to the nearby eigenvalues essentially deflates them and thus gives good convergence for Lan-DR.  For eigenvalue 30, Lan-DR trails unrestarted Lanczos, but is still competitive.  This is significant, since Lan-DR(100,40) requires storage of only about 100 vectors compared to nearly 3000 for unrestarted Lanczos.  


\begin{figure}
\includegraphics[width=4in]{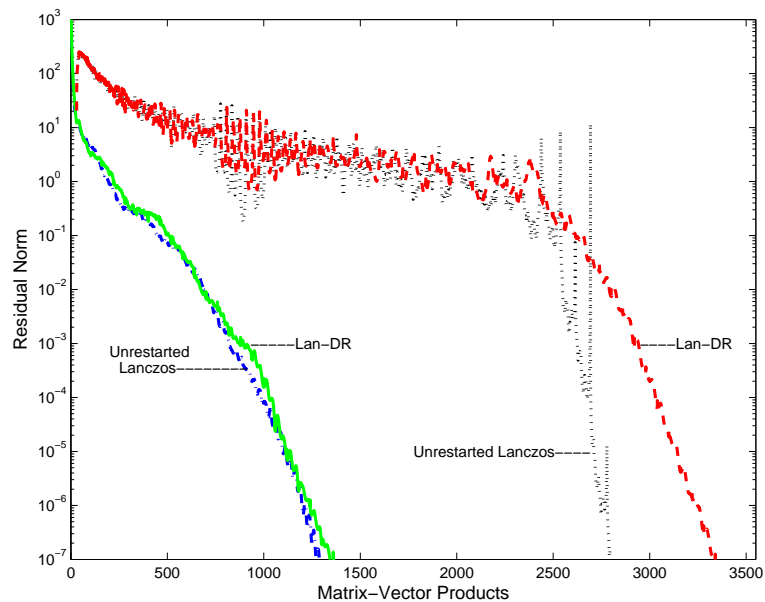}
\caption{Comparison of Lan-DR(100,40) with unrestarted Lanczos for first and 30th eigenvalues.}
\end{figure}

Next, we look at the solution of the linear equations.  Figure 3.3 has Lan-DR with three choices of $m$ and $k$ and also has CG.  The convergence of Lan-DR(100,40) is very close to that of CG.  The deflation of eigenvalues in Lan-DR(100,40) allows it to compete with an unrestarted method such as CG.  Lan-DR(100,10) is not too far behind CG, but Lan-DR(30,10) restarts too frequently and is much slower.  


\begin{figure}
\includegraphics[width=4in]{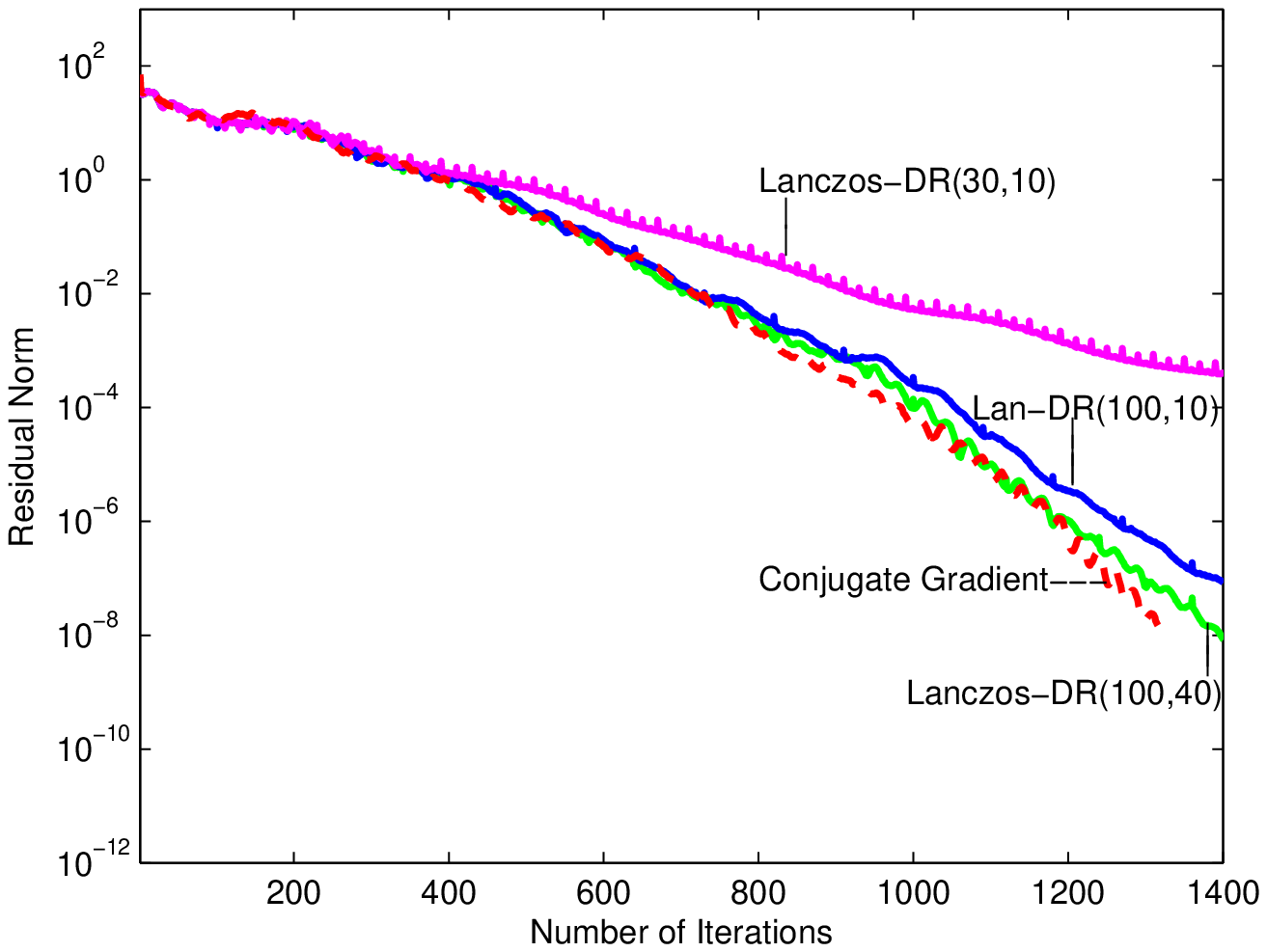}
\caption{Comparison of Lan-DR with CG for solving linear equations.}
\end{figure}


\section{Reorthogonalization}

It was mentioned earlier that some reorthogonalization is necessary to control roundoff error.   We now look at this in more detail and give several possible approaches.  The first is full reorthogonalization which takes every Lanczos vector formed by the three-term recurrence and reorthogonalizes it against every previous vector.  The expense for this in terms of vector operations of length $n$ varies from about $2k$ to $2m$ per iteration.  The second approach is periodic reorthogonalization~\cite{Gr81}.  For this, we always reorthogonalize the $v_{k+1}$ and $v_{k+2}$ vectors, then at regular intervals reorthogonalize two consecutive vectors (see~\cite{WuSi,Stv2} for why consecutive vectors need to be reorthogonalized).  The cost for this varies as with full reorthogonalization when it is applied, but it saves considerably if the reorthogonalization is not needed frequently.  Next is partial reorthogonalization (PRO)~\cite{Si84,Si84b,WuSi,Stv2} which monitors loss of orthogonality and thus determines when to reorthogonalize.  As suggested in~\cite{WuSi}, we use ``global orthogonalization", so when we reorthogonalize, it is against all previous vectors.  As with periodic, we always reorthogonalize the $v_{k+1}$ and $v_{k+2}$ vectors and always reorthogonalize two consecutive vectors.  This approach can be cheaper than periodic reorthogonalization, because it waits until reorthogonalization is needed.  

The next three reorthogonalization methods are related to the three above, but reorthogonalization is done only against the first $k$ vectors.  Here we are using the idea from selective reorthogonalization (SO)~\cite{PaSc} that orthogonality is only lost in the direction of converged or converging Ritz vectors~\cite{Paige71,Pa}.  Since restarting is used, normally only the $k$ Ritz vectors that are kept at the restart have an opportunity to converge.  There can be exceptions as will be discussed in Examples 3 and 4 below.  The first SO-type approach has been used in the previous section.  It is called k-SO and has reorthogonalization at every iteration against the $k$ Ritz vectors.  This requires about $2k$ vector operations per iteration versus an average of about $k+m$ vector operations for full reorthogonalization.  The last two methods are k-periodic and k-PRO.  These are the same as periodic and PRO, except they reorthogonalize only against the first $k$ vectors. 

{\it Example 2.} 
We consider again Lan-DR for the matrix of Example 1.  For this problem, loss of orthogonality is controlled by the restarting and the reorthogonalization of the $v_{k+1}$ and $v_{k+2}$ vectors at the restart.  No further reorthogonalization is needed.  Table 4.1 show a comparison with full reorthogonalization.  The second column gives the loss of orthogonality as measured by $\|V^{T}_{m}V_{m}-I_{m \times m}\|$ at the end of 57 cycles.  The next two columns have the residual norms of the first and thirtieth Ritz pairs.  While full reorthogonalization gives greater orthogonality of the Lanczos vectors, the Ritz vectors end up with similar accuracy.  The thirtieth eigenvector continues to converge beyond cycle 57 and eventually reaches residual norm of $3.7 \times 10^{-12}$ even with the approach of reorthogonalizing only at the restart.  For this example, the Ritz vectors converge slowly enough that we don't have a Ritz vector appear and significantly converge in one cycle (see Figure 3.1).  So before a eigenvector has converged, an approximation to it is among the group of Ritz vectors that $v_{k+1}$ and $v_{k+2}$ are reorthogonalized against.  This explains why reorthogonalizing at restarts turns out to be often enough.

This example points out that restarting can make reorthogonalization easier.  Reorthogonalization against only 40 vectors is done for two vectors every 60 iterations.  If we compare to unrestarted Lanczos using PRO with global reorthogonalization and with tolerance on loss of orthogonality of square root of machine epsilon, the number of vectors reorthogonalized is similar.  However, as the unrestarted Lanczos iteration proceeds, there are many previous vectors to reorthogonalize against.  Also unrestarted Lanczos with PRO gives converged eigevectors with residual norms of just below $10^{-6}$ compared to well below $10^{-11}$ for Lan-DR(100,40) with reorthogonalization only at the restart.  We note however, that the PRO tolerance can be adjusted for more frequent reorthogonalization and greater accuracy.  

The next matrix is designed so that Lan-DR needs more reorthogonalization.  We compare approaches and look at some potential problems.

\begin{table}

\caption{Full reorthogonalization vs. reorthogonalize only at the restart}

\begin{center}
\begin{tabular}{|c|c|c|c|c|}  \hline\hline
       & orthogonality  & rn 1  & rn 30   \\ \hline
\hline
k+1, k+2 vectors only  &  $2.2 \times 10^{-12}$ & $5.6 \times 10^{-12}$ & $9.9 \times 10^{-9}$   \\ \hline
full reorthog. &  $1.2 \times 10^{-14}$ & $6.7 \times 10^{-12}$ & $9.9 \times 10^{-9}$  
\\ \hline \hline

\end{tabular}
\end{center}

\end{table}

{\it Example 3.} 
Let the matrix be diagonal with dimension $n = 5000$ and diagonal elements $1, 2, 3, \ldots 9, 10,$ $100, 101, 102, \ldots , 5088, 5089.$  The right-hand side is a random normal vector.  We use 12 cycles of Lan-DR(120,40) with the reorthogonalization approaches described at the beginning of this section.  More reorthogonalization is needed than in Example 2, because eigenvectors converge quicker.  Table 4.2 has the results.  Two different tolerances on the loss of orthogonality for the PRO methods are used, square root of machine epsilon and three-quarters power.  The second column of the table gives the frequency of reorthogonalization (of two vectors) for periodic methods.  The next three columns are the same as columns in the previous table.  The last column has another measure of the effect of the roundoff error.  It gives the number of iterations needed to solve a second right-hand side using the deflated CG method which will be given in the next section.  We see that k-periodic reorthogonalizing of two vectors every 40 iterations gives fairly good results.  Accuracy drops as the reorthogonalization is done less frequently.  With frequency of 75, Lan-DR is still able to compute some eigenvectors accurately, but unlike with frequency of 70, multiple copies of eigenvalues appear.  Also, Lan-DR is not longer helpful for the solution of the second right-hand side.  Partial reorthogonalization gives results somewhat similar to periodic restarting without the need to select the frequency ahead of time.  For k-PRO with $\epsilon_0^{.75}$, a total of 82 vectors are reorthogonalized and with $\epsilon_0^{.5}$, there are 44 reorthogonalizations.  This compares to 50 for k-periodic with frequency of 40.    

\begin{table}

\caption{Compare reorthogonalization methods}

\begin{center}
\begin{tabular}{|c|c|c|c|c|c|}  \hline\hline
       & reor. freq.  & orthogonality  & rn 1  & rn 30  & 2nd rhs it's \\ \hline
\hline
full       & 1 &  $1.3 \times 10^{-14}$ & $7.5 \times 10^{-12}$ & $1.7 \times 10^{-10}$ & 57  \\ \hline
k-SO       & 1 &  $1.8 \times 10^{-11}$ & $7.5 \times 10^{-12}$ & $1.8 \times 10^{-10}$ & 57  \\ \hline
k-periodic & 40 & $6.7 \times 10^{-10}$ & $5.8 \times 10^{-12}$ & $1.2 \times 10^{-8}$ & 57  \\ \hline
           & 60 & $1.7 \times 10^{-7}$  & $4.8 \times 10^{-12}$ & $6.6 \times 10^{-7}$ & 57  \\ \hline
           & 70 & $4.0 \times 10^{-5}$  & $5.9 \times 10^{-12}$ & $3.7 \times 10^{-5}$ & 70  \\ \hline
           & 75 & $9.0 			 $  & $4.6 \times 10^{-12}$ & -		         & 218 \\ \hline
          
periodic   & 40 & $3.1 \times 10^{-10}$  & $5.8 \times 10^{-12}$ & $2.0 \times 10^{-9}$ & 57  \\ \hline
	     & 70 & $2.8 \times 10^{-6}$  & $5.9 \times 10^{-12}$ & $2.0 \times 10^{-5}$ & 57  \\ \hline
 	     & 75 & $2.4 \times 10^{-5}$  & $4.6 \times 10^{-12}$ & $5.5 \times 10^{-5}$ & 57  \\ \hline

           & 80 & $5.5               $  & $5.5 \times 10^{-12}$ & $8.0 \times 10^{-3}$ & 213 \\ \hline
k-PRO, $\epsilon_0^{.5}$   & - & $8.4 \times 10^{-9}$  & $4.9 \times 10^{-12}$ & $1.5 \times 10^{-7}$ & 57  \\ \hline
k-PRO, $\epsilon_0^{.75}$  & - & $1.0 \times 10^{-11}$ & $4.9 \times 10^{-12}$ & $2.1 \times 10^{-10}$ & 57 \\ \hline
PRO, $\epsilon_0^{.5}$     & - & $7.6 \times 10^{-9}$  & $4.9 \times 10^{-12}$ & $1.1 \times 10^{-7}$ & 57  \\ \hline
PRO, $\epsilon_0^{.75}$    & - & $4.8 \times 10^{-12}$ & $4.9 \times 10^{-12}$ & $1.7 \times 10^{-10}$ & 57 

\\ \hline \hline

\end{tabular}
\end{center}


\end{table}

{\it Example 4.}
We continue with same matrix.
Because it has fairly rapidly converging eigenvalues, it can be used to demonstrate a possible problem with reorthogonalizing against only the $k$ saved Ritz vectors.  Table 4.3 has results for Lan-DR with k-SO for $k=40$ and changing values of $m$.  Each test is run the number of cycles so that the total number of matrix-vector products is 1000 or just over.  Even though k-SO reorthogonalizes at every iteration, there is increasing loss of orthogonality as $m$ increases, until a high level of orthogonality comes back at $m=200$.  We explain first what happens when $m=140$.  Recall that this matrix has 10 well separated eigenvalues.  Also recall we fully reorthogonalize in the first cycle.  After that first cycle of Lan-DR(140,40) with k-SO, there are only seven Ritz values below 10.  After the second cycle, all ten small eigenvalues have converged to a high degree.  Some orthogonality is lost in that cycle, because by the end of the cycle, there are converged Ritz vectors in the subspace that are not reorthogonized against.  The effect is more extreme with $m=180$, because then nine of the ten small eigenvalues converge part way in the first cycle while one eigenvalue is missing (the one at 7.0).  That missing one converges very early in the second cycle and orthogonality is lost after that.  So it is dangerous to use k-SO if some eigenvectors converge very rapidly (within one cycle).  On the other hand, k-SO and the other k-methods have been successful for other problems, such as in Example 2 and for the large QCD matrices that are in later experiments.  The next example shows another possible problem with k-SO.

\begin{table}

\caption{An effect of rapidly converging eigenvectors.}

\begin{center}
\begin{tabular}{|c|c|c||c|c|}  \hline\hline

 	      & k-SO  &   & full reorthogonalization  &     \\ \hline
 m , k      & orthogonality  & rn30  & orthogonality  & rn 30    \\ \hline
\hline
100, 40      & $4.8 \times 10^{-12}$ & $1.2 \times 10^{-10}$ & $1.2 \times 10^{-14}$ & $6.3 \times 10^{-11}$   \\ \hline

120, 40      & $1.8 \times 10^{-11}$ & $1.8 \times 10^{-10}$ & $7.5 \times 10^{-12}$ & $1.8 \times 10^{-10}$   \\ \hline
140, 40      & $1.1 \times 10^{-8}$  & $3.6 \times 10^{-7}$  & $5.8 \times 10^{-12}$ & $1.2 \times 10^{-8}$    \\ \hline
160, 40      & $1.3 \times 10^{-3}$  & $6.3 \times 10^{-5}$  & $4.8 \times 10^{-12}$ & $6.6 \times 10^{-7}$    \\ \hline
180, 40      & 1.0	             & $1.4 $     		 & $5.9 \times 10^{-12}$ & $3.7 \times 10^{-5}$    \\ \hline
200, 40      & $1.2 \times 10^{-12}$ & $3.4 \times 10^{-11}$ & $4.6 \times 10^{-12}$ & $3.4 \times 10^{-11}$		         
       
\\ \hline \hline

\end{tabular}
\end{center}
\end{table}

{\it Example 5.} 
For matrices with outstanding eigenvalues other than the small ones that can converge in one cycle, care must be taken.  For k-SO to work, Ritz vectors corresponding to those eigenvalues need to be included in the Ritz vectors that are saved at the restart.  We use the same matrix as in the previous example, except the largest eigenvalue is changed from 5089 to 5400.  This eigenvalue is now outstanding enough to converge rapidly.  Lan-DR(120,40) with k-SO loses orthogonality of its basis.  After 12 cycles, the orthogonality level is $3.4 \times 10^{-3}$, and it is actually worse (near 1) at earlier cycles.  As mentioned, this can be fixed by including the large eigenpair among those selected to be saved for the next cycle.  Another option is to use the reorthogonalization against all previous vectors instead of just the first $k$.

\section{Multiple Right-hand Sides}

Next, solution of systems with multiple right-hand sides is considered.  We suggest a simple approach that uses the eigenvectors generated during the solution of the first right-hand side to deflate eigenvalues from the solution of the second right-hand side.  First a projection is done over the Ritz vectors at the end of Lan-DR for the first right-hand side.  Then the standard conjugate gradient method is used.  We call this approach deflated CG or D-CG, since it deflates out eigenvalues before applying CG.  It is similar to the init-CG approach~\cite{ErGu} that also does a projection before CG.  However, init-CG projects over the entire Krylov subpace generated by CG while solving the first right-hand side, while D-CG uses a projection over a compact space that has the important eigen-information.

\vspace{.10in}
\begin{center}
\textbf{D-CG}
\end{center}
\begin{enumerate}
 \item After applying the initial guess $\tilde x_0$, let the system
of equations be $A(x-\tilde x_0) = r_0$.  
 \item If it is known that the right-hand sides are closely related, project over the previous computed solution vectors.
 \item Apply the Galerkin Projection for $V_k$.  This uses the $V_{k+1}$ and $\overline T_k$ matrices from (\ref{recur1}) that were developed while solving the first right-hand side with Lan-DR.  Specifically, solve $T_k d = V_k^T r_0$, where $T_k$ is the diagonal $k$ by $k$ portion of $\overline T_k$, and let the new approximate solution be $\tilde x = \tilde x_0 + V_k d$ and the new residual be $r = r_0 - A V_k d = r_0 - V_{k+1} \overline T_{k} d$.
 \item Apply CG until satisfied with convergence.
\end{enumerate} 
\vspace{.15in}



\subsection{Examples}

\ 

{\it Example 6.} We consider the same matrix as in Example 1 and solution of a second right-hand side.  The first right-hand side system has been solved with Lan-DR(100,40).  We first illustrate how increasing the accuracy of the approximate eigenvectors helps the convergence of the second right hand side.  Figure 5.1 has convergence curves for the second right-hand side with D-CG when Lan-DR has been run different numbers of cycles.   CG is also given and it lags behind all the D-CG curves.  The Lan-DR linear equations relative residual converges at 20 cycles.  However, D-CG is not as effective as it can be if Lan-DR has only been run 20 cycles.  The eigenvectors are not all accurate enough to deflate out the eigencomponents from the residual of the second right-hand side, and eventually CG has to deal with these components and this slows convergence.
D-CG after 40 cycles converges rapidly and does not slow down as it procedes.  Using 60 cycles of Lan-DR is only a little better.  Figure 5.2 has the components in the directions of the eigenvectors corresponding to the 80 smallest eigenvalues of the residual vector for D-CG on the second right-hand side after the projection over the approximate eigenvectors that come fromt the first right-hand side.  These components improve significantly as Lan-DR on the first is run more cycles.  

\begin{figure}
\includegraphics[width=4in]{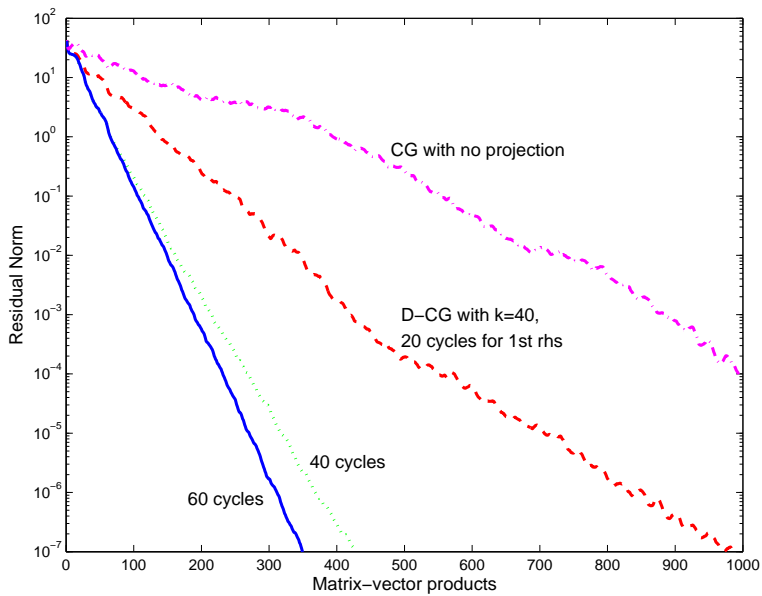}
\caption{Solving the first right-hand side to different levels of accuracy.}
\end{figure}

\begin{figure}
\includegraphics[width=4in]{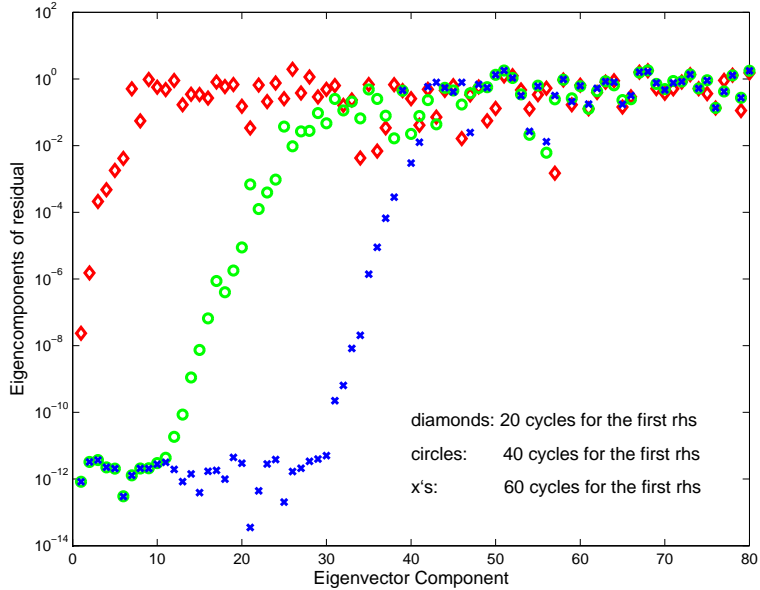}
\caption{Eigencompents for residual for the second right-hand side after solving the first right-hand side to different levels of accuracy.}
\end{figure}

We next consider varying the number of approximate eigenvectors that are used for deflation.  For the first right-hand side, Lan-DR(m,k) is run for 50 cycles with changing $k$ and with $m=k+60$.  Figure 5.3 has the convergence results for then applying D-CG to the second right-hand side.  With $k=10$ eigenvectors, D-CG is already significantly better than regular CG, but deflating more eigenvalues is even better.  For this example there is a significant jump upon going from 80 to 120 eigenvectors.  This happens because having 120 gets past the 100 clustered eigenvalues and pushes well into the rest of the spectrum.  

\begin{figure}
\includegraphics[width=4in]{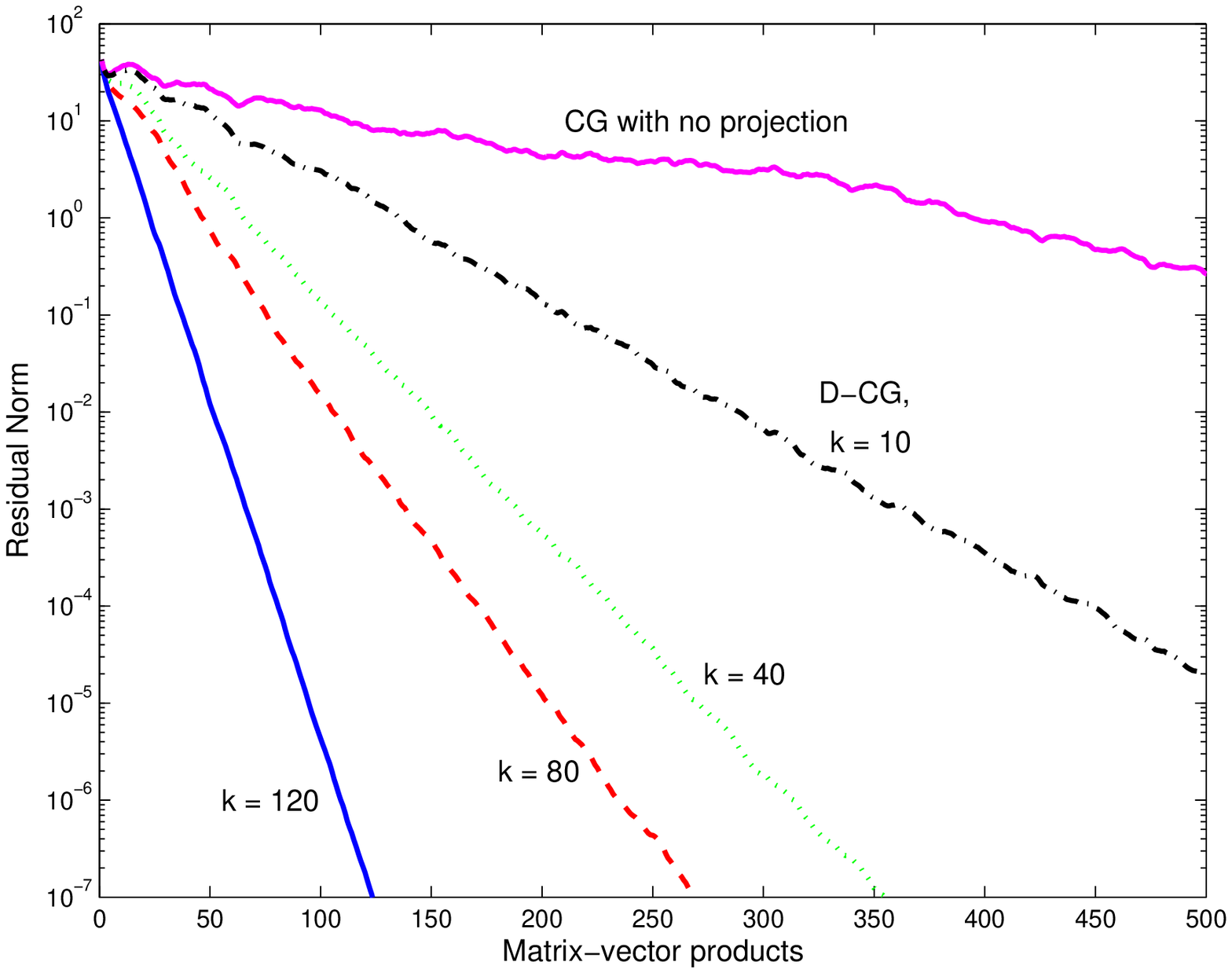}
\caption{Deflating different numbers of eigenvalues.}
\end{figure}


Figure 5.4 shows that Lan-DR/D-CG can come out ahead of regular CG in terms of matrix-vector products even if we spend more on Lan-DR for the first right-hand side.  We use 10 right-hand sides.  The first is solved with 44 cycles of Lan-DR(180j,120).  Then the other nine use D-CG with relative residual tolerance of $10^{-8}$.  This all takes about as many matrix-vector products as solving three systems with CG.

\begin{figure}
\includegraphics[width=4in]{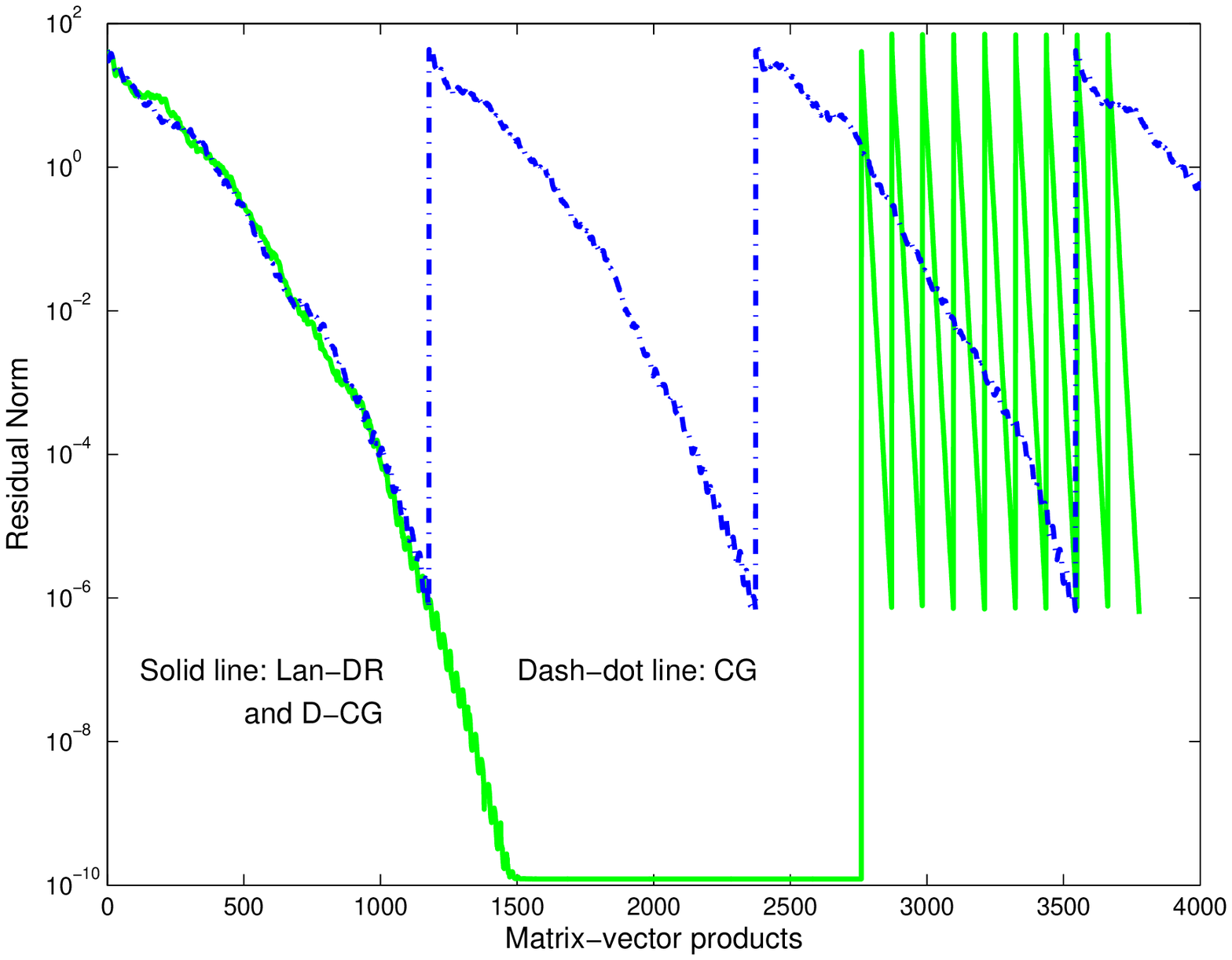}
\caption{Lan-DR and D-CG versus regular CG for 10 right-hand sides.}
\end{figure}

\subsection{Comparison with block-CG}

Block-CG~\cite{OL80,NiYe,Scm04} generates a Krylov subspace with each right-hand side as starting vector then combines them all together into one large subspace.  This large subspace can generally develop approximations to the eigenvectors corresponding to the small eigenvalues, so block-CG naturally deflates eigenvalues as it goes along.  As a result, block-CG can be very efficient in terms of matrix-vector products.  

Block methods require all the right-hand sides be available at the same time, while Lan-DR/D-CG only needs one right-hand side at a time.  Block methods have extra orthogonalization expense compared to non-block methods.  Simple block-CG can be unstable, particularly if the right-hand sides are related to each other.  This can be controlled by the somewhat complicated process of removing (``deflating") right-hand sides~\cite{NiYe}.  

As mentioned, block-CG can converge quickly.  For instance with the matrix from Example 1 and 20 random right-hand sides, block-CG needs only 3100 matrix-vector products.  Lan-DR(180,120)/D-CG uses 4909.  The next example shows that Lan-DR/D-CG can be competitive with block-CG.

\begin{figure}
\includegraphics[width=4in]{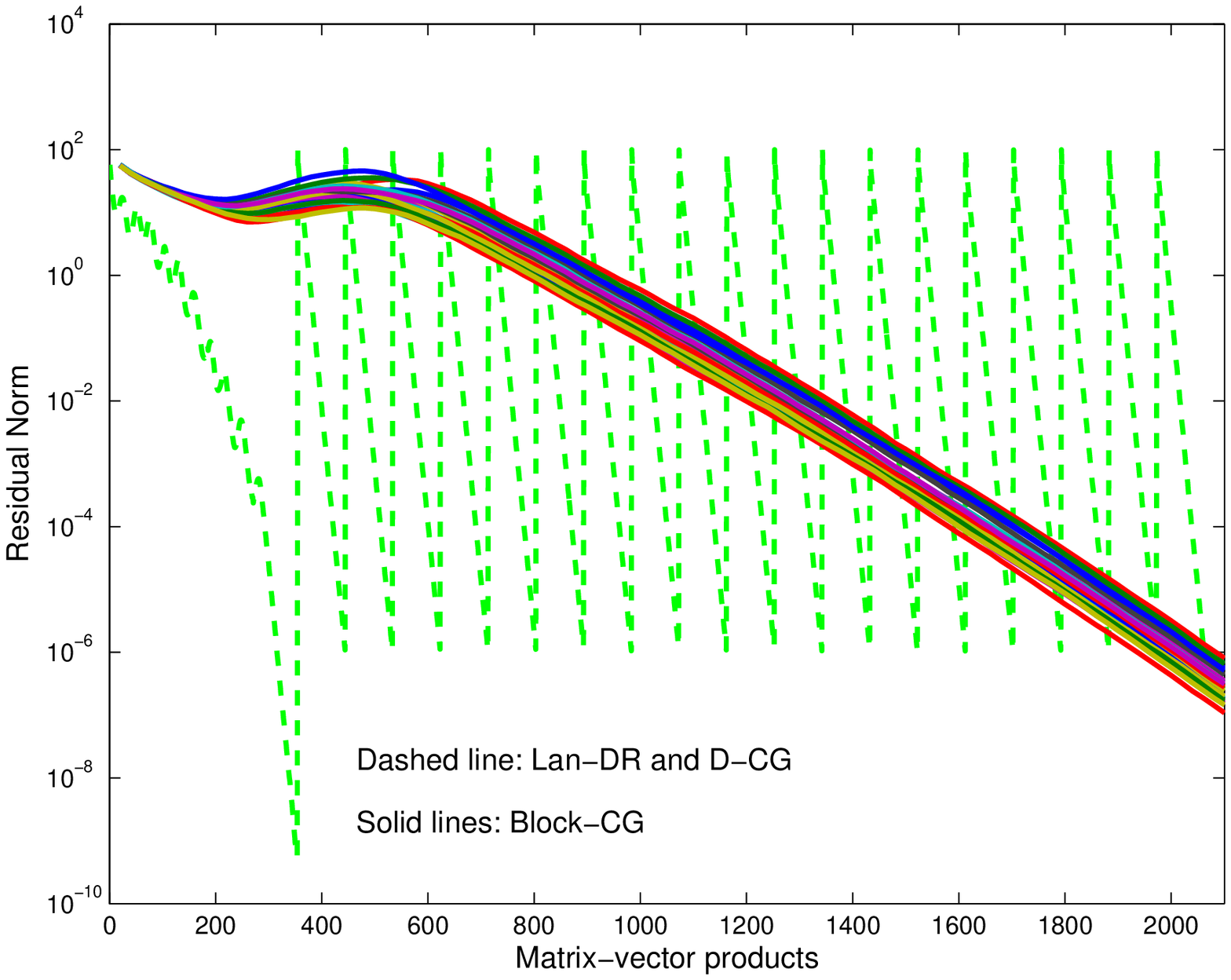}
\caption{Comparison with Block-CG for 20 right-hand sides.}
\end{figure}

{\it Example 7.} We use the same matrix as in Example 3, but with $n=10000$ (so the largest element is 10089).  We compare the Lan-DR/D-CG approach with block-CG for 20 random right-hand sides.  Lan-DR has $m=100$ and $k=15$ and runs for four cycles.  Figure 5.5 shows that the two approaches converge at almost the same number of matrix-vector products.  However, Lan-DR has less orthogonalization expense.  Lan-DR with k-periodic reorthogonalization of two vectors every 40 iterations followed by D-CG for the remaining 19 right-hand sides uses 14,000 vector operations of length $n$.  Block-CG needs 177,000.

\subsection{Related right-hand sides}


\ 

{\it Example 8.}  We use the same matrix as in the previous example.  There are again 20 right-hand sides, but this time the first is chosen randomly and the others are chosen as $b^{(i)} = b^{(1)} + 10^{-3} * ran^{(i)}$, where $ran^{(i)}$ is a random vector (elements chosen randomly with Normal(0,1) distribution.  The convergence tolerance is moved to relative residual below $10^{-6}$, because block-CG has instability after that point.  Figure 5.6 shows that Lan-DR does a better job of taking advantage of the related right-hand sides.  However, as mentioned earlier, block-CG can be improved by removing right-hand sides once they become linearly dependent.  

\begin{figure}
\includegraphics[width=4in]{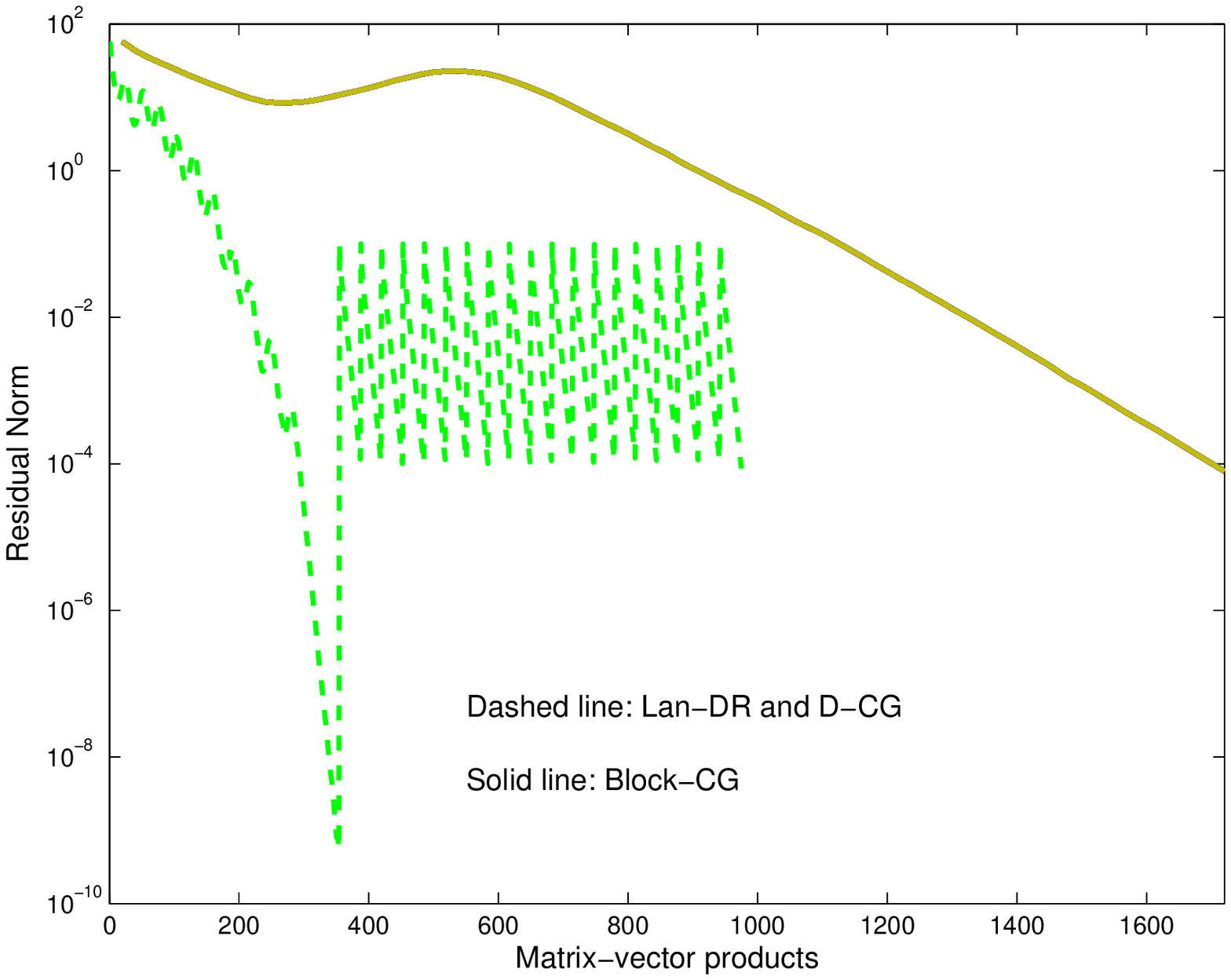}
\caption{Comparison with Block-CG for 20 related right-hand sides.}
\end{figure}

\subsection{Example from QCD}

Many problems in lattice quantum chromodynamics (lattice QCD) have large systems of equations with multiple right-hand sides.  For example, the Wilson-Dirac formulation~\cite{Frommer,Qcdconf2} and overlap fermion~\cite{NaNe,FrOverlap2} computations both lead to such problems.  Deflation of eigenvalues is used for QCD problems, for example in~\cite{dF,EdHeNa,DoLeLiZh,NeEiLiNeSc,Qcdconf2,gproj,StOr,Lusch}.  QCD matrices have complex entries.  They generally are non-Hermitian, but it may be possible to change into Hermitian form.  This can be done by multiplying the system by $\gamma_5$~\cite{Frommer} or multiplying by the complex conjugate of the QCD matrix.  We consider the first of these options, which gives an indefinite matrix.  

{\it Example 9.} 
We choose a large QCD matrix of size $n = 1.5$ million.  The $\kappa$ value is set near to $\kappa$-critical, which makes it a difficult problem.  There are at least a dozen right-hand sides for each matrix and sometimes over a hundred.  The first right-hand side is solved using Lan-DR(m,k) with several values of $k$.  The linear equations solution does not converge past a residual norm of 0.036.  This shows Lan-DR may not be stable for an indefinite problem and helps motivate the development of Minres methods in the next section.  However, Lan-DR does generate useful eigenvectors that can be used to solve the other right-hand sides.  Figure 5.7 shows the convergence for solution of the second right-hand side system using D-CG with three choices of $k$.  These are compared to CG.  We note that deflating 20 eigenvalues gives a big improvement over regular CG.  Using 150 eigenvectors is almost an order of magnitude improvement over CG.

\begin{figure}
\includegraphics[width=4in]{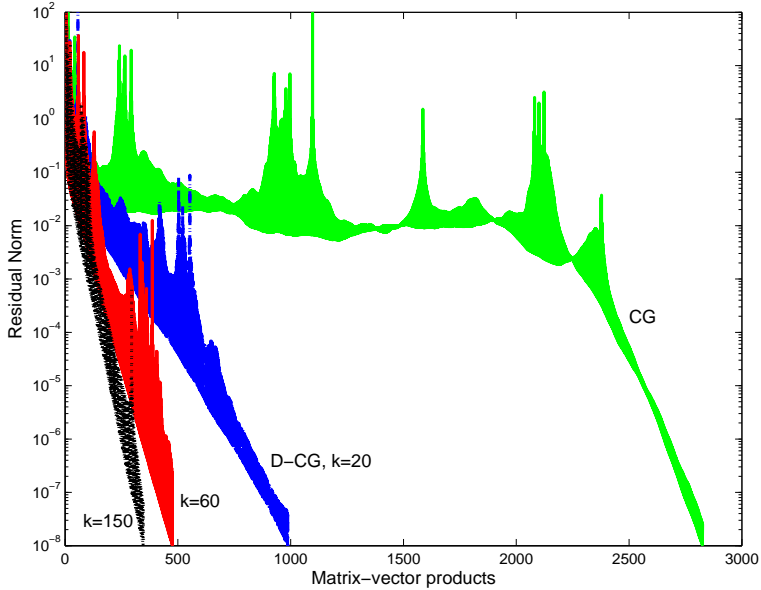}
\caption{Comparison of CG to D-CG with varying numbers of eigenvalues deflated for a large QCD matrix.}
\end{figure}

The results in Figure 5.7 are fairly typical.  Figure 5.8 shows the even iterations only for five configurations (matrices).  D-CG with $k = 100$ eigenvalues deflated is compared with CG.  There is some variance in CG, but with 100 small eigenvalues taken out, the convergence of D-CG is almost identical for all matrices.  

\begin{figure}
\includegraphics[width=4in]{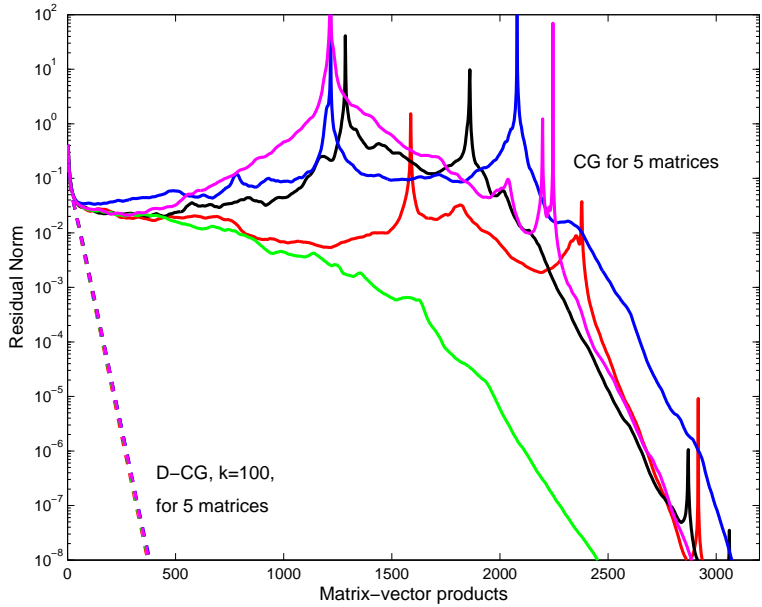}
\caption{Comparison of CG to D-CG with k=100 eigenvalues deflated for five QCD matrics.}
\end{figure}

\section{Minres-DR and deflated Minres}

We now consider symmetric/Hermitian indefinite problems.  For stability, we need minimum residual~\cite{PaSa,Sa84,Sa96} versions of our methods.  Instead of Lan-DR, a method Minres-DR can be used.  It solves the linear equations problem with a minimum residual projection and it computes harmonic Ritz vectors~\cite{IE,Fr92,PaPavdV,IEN,Stv2,HRAM} instead of regular Ritz vectors.  Harmonic Ritz approximations are more reliable for interior eigenvalues.  We next give the steps only that are changed from Lan-DR to Minres-DR.  

\vspace{.10in}
\begin{center}
\textbf{Minres-DR(m,k)}
\end{center}
\begin{itemize}

\item[3.] \textit{Eigenvector computation.}  Compute the $k$
smallest (or others, if desired) harmonic Ritz values, $\theta_{i}$, and let $g_{i})$ be the corresponding vectors.  See~\cite{HRAM} for more, including residual norm formulas.

\item[4.] \textit{Linear equations.}  Let $c_{m+1} = V_{k+1} r_0$.  Solve the least squares problem $min ||c_{m+1} - \overline T_m d||$ for $d$ and set $\tilde x = x_0 + V_m d$.    Then $r = r_0 - A \tilde x = r_0 - V_{m+1} \overline{T}_{m} d$. If satisfied with convergence of the linear equations and the eigenvalues, can stop.  If not, let the new $x_0 = \tilde x$ and $r_0 = r$ and continue.   

\item[5.] \textit{Restart.} Let $P$ be the $m+1$ by $k+1$ matrix whose first $k$ columns come from orthonormalizing the $g_i$ vectors (and adding zeros for the $m+1$ row).  Let $e_m$ be the $m$th coordinate vector of length $m$.  Then the $k+1$ column of $P$ is the vector $[-t_{m+1,m} T_m^{-T}e_m, 1]^T$~\cite{RoFi} orthonormalized against the previous columns.  The new $V$ and $T$ matrices are formed from the old ones: $V_{k+1}^{new} = V_{m+1}P$ and $\overline{T}_{k}^{new} = P^{T} \overline T_m P_{m,k}$, where $P_{m,k}$ has the first $m$ columns and $k$ rows of $P$.
Set $V_{k+1} = V_{k+1}^{new}$ and $\overline{T}_{k} = \overline{T}_{k}^{new}$. 
 
\end{itemize}
\vspace{.15in}

For the second and subsequent right-hand sides, D-Minres (deflated Minres) can be used.  Like D-CG, it has a projection over the approximate eigenvectors, but this is followed by Minres~\cite{PaSa}.

{\it Example 10.}  For an indefinite matrix, we use a diagonal matrix of dimension $n=1000$ whos diagonal entries are generated with random numbers distributed Normal(0,1) that are shifted 2.0 to the right.  Then there are 22 negatives among the 1000 eigenvalues.  The five closest to the origin are -0.015, -0.033, 0.041, -0.53, and -0.57.  Indefinite problems can be difficult if there are very many eigenvalues on both sides of the origin, unless the eigenvalues are well separated from the origin.  Figure 6.1 has a comparison of the Minres methods with the Galerkin methods Lan-DR and D-CG for solution of three right-hand sides.  For the second and third right-hand sides, D-Minres is used for both tests because the Matlab CG function found instabity and would not proceed.  We see that the results are fairly similar, except Minres-DR converges more smoothly than Lan-DR.  It seems that stability concerns are the main reason to use the Minres versions.


\begin{figure}
\includegraphics[width=4in]{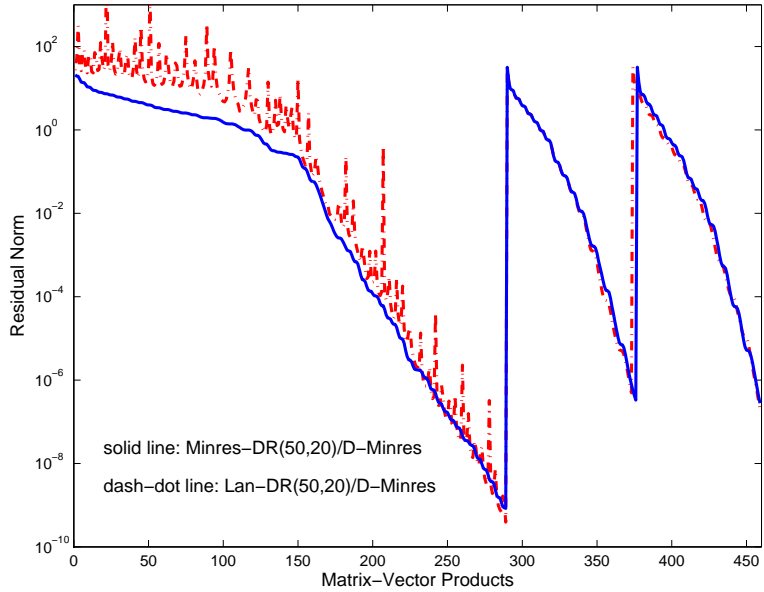}
\caption{Minres-DR vs. Lan-DR for the first rhs with D-Minres for the 2nd and 3rd rhs's.}
\end{figure}

{\it Example 11.}  The Minres methods are next applied to the first large QCD matrix from Example 9.  Figure 6.2 shows Minres-DR(200,60) as diamonds at the end of each cycle.  Minres-DR converges almost as fast as standard unrestarted Minres.  We also see that D-Minres gives greater improvement as the number of approximate eigenvectors is increased.  Finally, D-Minres converges a little faster than D-CG does in Example 9.  For instance, following Lan-DR(200,150), D-CG takes 348 matrix-vector products.  Meanwhile, D-Minres needs only 310 when following Minres-DR(200,150).

\begin{figure}
\includegraphics[width=4in]{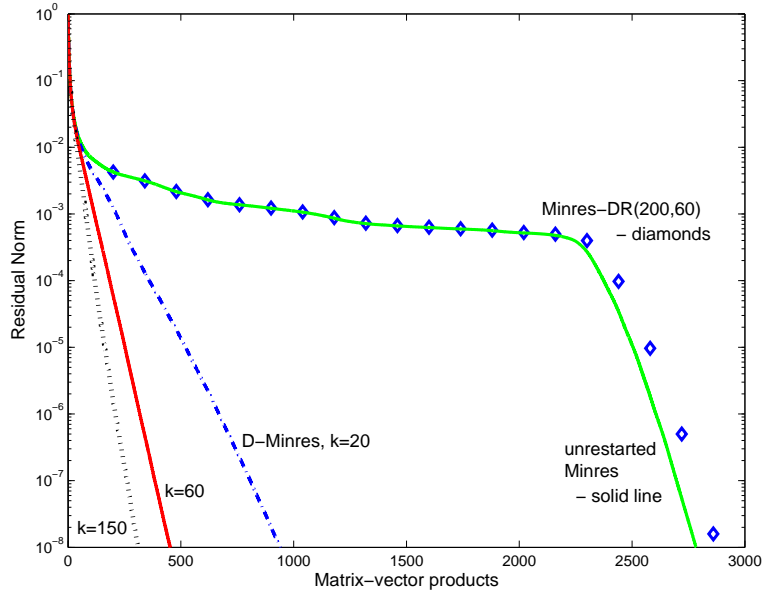}
\caption{Minres-DR and D-Minres for the large QCD matrix.}
\end{figure}

\section{Conclusion}

Lan-DR is a restarted Lanczos method that solves a symmetric/Hermitian positive definite system of linear equations and simultaneously computes eigenvalues and eigenvectors.  The restarting allows the computation of eigenvectors when there are limits on storage.  The presence of the approximate eigenvectors in the subspace helps the convergence.  In fact, the convergence is often close to that of unrestarted Lanczos for both the linear equations and the eigenvalues, in spite of the restarting.  

There are several options for reorthogonalizing.  Included are methods that only reorthogonalize against the $k$ saved Ritz vectors.  For these methods, there may be trouble if there are rapidly converging eigenvalues that converge in one cycle.  Restarting generally keeps reorthogonalization costs down.  

For subsequent right-hand sides, deflated CG first uses a projection over the eigenvectors that were computed by Lan-DR while solving the first right-hand side system, then applies CG.  For difficult problems, the convergence of CG can be much faster after the small eigenvalues are deflated out.  Experiments on large problems from QCD back this up.

Minres-DR is a version of Lan-DR for indefinite problems.  For indefinte systems with multiple right-hand sides, a deflated Minres method is given.



\bibliography{morgan}

\end{document}